\documentclass[reprint,aps,prl,showpacs]{revtex4-1}

\usepackage{graphicx}
\usepackage{bm}
\usepackage{hyperref}
\usepackage[latin1]{inputenc}
\usepackage{amsmath}
\usepackage{amsfonts}
\usepackage{amssymb}
\usepackage{amsthm}

\makeatletter
\newcommand{\colorcaption}[2][]{%
  \begingroup%
  \renewcommand{\@caption@fignum@sep}{ (Color online). }%
  \caption[#1]{#2}%
  \endgroup%
}
\makeatother

\begin{document}


\title{Laminar Chaos}

\author{David M\"uller}
 \email{david.mueller@physik.tu-chemnitz.de}
\author{Andreas Otto}%
 \email{otto.a@mail.de}
\author{G\"unter Radons}
 \email{radons@physik.tu-chemnitz.de}
\affiliation{Institute of Physics, Chemnitz University of Technology, 09107 Chemnitz, Germany}
%


\date{\today}

\begin{abstract}
We show that the output of systems with time-varying delay can exhibit a new kind of chaotic behavior characterized by laminar phases, which are periodically 
interrupted by irregular bursts. Within each laminar phase the output intensity remains almost constant, but its level varies chaotically from phase to 
phase. In scalar systems the periodic dynamics of the lengths and the chaotic dynamics of the intensity levels can be understood and also tuned via two 
one-dimensional maps, which can be deduced from the nonlinearity of the delay equation and from the delay variation, respectively.
\end{abstract}

\maketitle

Time-delay systems are known for their rich variety of dynamical behaviors \cite{2010just, 2013soriano,franz_changing_2007}. Especially systems with large delay are of general interest \cite{wolfrum_complex_2010} and exhibit interesting phenomena such as multistability and multiple chaotic attractors \cite{ikeda_high-dimensional_1987, 2009yanchuk, williams_synchronization_2013}. By introducing a space-time representation many aspects of spatially extended systems can also be found in time-delay systems \cite{yanchuk_spatio-temporal_2017}, which are, for example, spatio-temporal intermittency and defects \cite{giacomelli_defects_1994}, or concepts such as Eckhaus instabilities \cite{wolfrum_eckhaus_2006} and convective instabilities \cite{giacomelli_relationship_1996}. Whereas these aspects are well understood for systems with constant delays, much less is known for variable delays although the latter typically provide more realistic models. Time-varying delays further increase the complexity of the dynamics \cite{2007senthilkumar, 2009radons, 2015fischer, 2016lazarus}. Some useful properties for application to chaos communication can be found in \cite{kye_characteristics_2004, *2004kye, 2007ghosh}. On the other hand, a delay modulation can stabilize the system \cite{madruga_effect_2001, 2011otto, *2013otto}. An analytical approach is available for systems with a fast delay variation, where the variable delay can be approximated by a time-invariant distributed delay \cite{2008gjurchinovski, *2010gjurchinovski, *2012juengling, *2014gjurchinovski, *2015sugitani, *grigorieva_stability_2015}. However, in general, this approximation is not valid. 

In this Letter, we demonstrate that introducing a time-varying delay can lead to hitherto unknown chaotic behavior. We characterize the new chaotic behavior by using the concept of dissipative delays, which was recently introduced in Refs.~\cite{otto_universal_2017, *muller_dynamical_2017}. Let us consider scalar delay differential equations (DDE) of the form
\begin{equation}
\label{eqn:sysdef}
\frac{1}{T} \dot{z}(t) = -z(t) + f(z(R(t))), \quad \text{with } R(t) = t-\tau(t).
\end{equation}
$R(t)$ is the retarded argument and $\tau(t)$ is the time-varying delay. Well-known systems with a structure as in Eq.~\eqref{eqn:sysdef} are the Ikeda equation, with $f(z)=\mu\,\sin(z)$ \cite{ikeda_multiple-valued_1979, ikeda_optical_1980}, describing the dynamics of an optical ring cavity with a nonlinear dielectric medium, and the Mackey-Glass equation, with $f(z)=\mu\,z/(1+z^{10})$ \cite{mackey_oscillation_1977}, which is a model for blood cell production. The DDE where $f(z)$ is given by the logistic map, i.e. $f(z)=\mu\,z(1-z)$, is an appropriate prototype system for deriving general properties of solutions of Eq.~\eqref{eqn:sysdef} \cite{adhikari_periodic_2008}, since the dynamics of the logistic map is well-understood. Systems described by Eq.~\eqref{eqn:sysdef} are interesting for many applications such as random number generators \cite{2008uchida, *2009reidler, *2010kanter}, chaos communication \cite{kye_characteristics_2004, *2004kye, 2007ghosh, goedgebuer_optical_1998, *vanwiggeren_optical_1998, *udaltsov_communicating_2001, *keuninckx_encryption_2017} or reservoir computing \cite{appeltant_information_2011, *larger_high-speed_2017} because they can be realized easily by optical, electronic, and optoelectronic setups. Often the parameter $T$ in Eq.~\eqref{eqn:sysdef} is large. For example, by rescaling time it can be seen that Eq.~\eqref{eqn:sysdef} with large $T$ is equivalent to a large delay. For systems with constant delay the large delay limit of DDEs has been extensively analyzed in the literature. For example, phenomena such as slowly oscillating periodic solutions \cite{chow_singularly_1983,mallet-paret_global_1986,adhikari_periodic_2008}, multistability of periodic solutions \cite{2009yanchuk, ikeda_successive_1982,mensour_chaos_1998, amil_organization_2015} and the scaling behavior of the Lyapunov exponents \cite{2011lichtner, 2011heiligenthal, *2013huys, *2015juengling} have been studied. For systems with large time-varying delay only few results are available. Some general aspects can be derived from the theory of singularly perturbed systems with state dependent delay \cite{mallet-paret_boundary_1992}. Moreover, the systems which were analyzed in Refs. \cite{grigorieva_quasi-periodic_2017, *grigorieva_discrete_2017} correspond to this class of systems.

\begin{figure}[t]
\raggedright
\hspace{5mm} a) \hspace{22mm} Turbulent chaos\\
\centering
\includegraphics[width=0.4\textwidth]{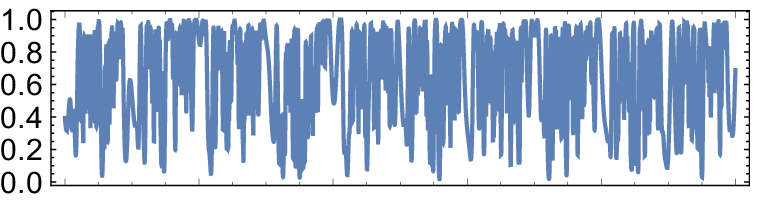}\\
\raggedright
\hspace{5mm} b) \hspace{22mm} Laminar chaos\\
\centering
\includegraphics[width=0.4\textwidth]{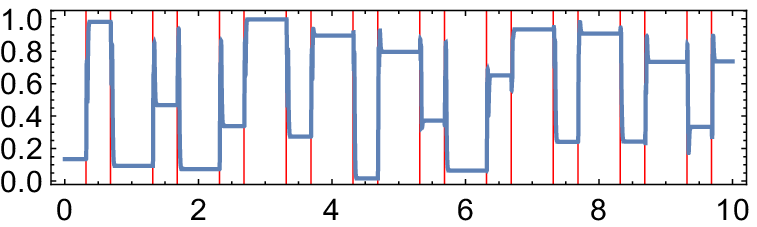}\\
\colorcaption{\label{fig:traj} Exemplary trajectories of DDE~\eqref{eqn:sysdef} with $f(z)=4\, z(1-z)$ and sinusoidal delay 
$\tau(t)=\tau_0+A\sin(2\pi\,t)$ for a) conservative delay ($\tau_0=1.54$) leading to turbulent chaos known from systems with constant delay \cite{ikeda_optical_1980} and for b) dissipative delay ($\tau_0=1.50$) leading to laminar chaos characterized by constant laminar phases between the attractive periodic points of the map $R^{-1}(t)\mod 1$ (vertical lines) and burst-like transitions between them ($T=200$, $A=0.9/(2\pi)$).}
\end{figure}

As demonstrated in Ref.~\cite{otto_universal_2017,muller_dynamical_2017} there are two classes of time-varying delays, leading to a fundamentally different tangent space dynamics. Systems with \emph{conservative delay} are equivalent to systems with constant delay, where 'equivalent' means that the systems are connected by an invertible timescale transformation $\varphi=\Phi(t)$ which leaves the dynamical quantities invariant. Systems with \emph{dissipative delay} cannot be mapped to systems with constant delay. Roughly speaking, for conservative delays the associated \emph{access map} $t'=R(t)$ is topological conjugate to the access map $\varphi'=R_c(\varphi):=\varphi-c$ of a system with constant delay $c$, that is, $\Phi(R(t))=R_c(\Phi(t))$. In contrast, for dissipative delays the access map $R$ exhibits mode-locking and no topological conjugacy to a map $R_c$ can be found. In other words, for dissipative delays the access map $R$ is dissipative, not to be confused with the dissipative nature of Eq.~\eqref{eqn:sysdef}, which holds for conservative as well as for dissipative delays. In this work we will demonstrate that under certain conditions a new kind of chaotic behavior can be found in systems with dissipative delay, which fundamentally differs from the known behavior for constant or time-varying conservative delay. In Fig.~\ref{fig:traj} the difference is illustrated by two exemplary chaotic trajectories of DDE~\eqref{eqn:sysdef}. Fig.~\ref{fig:traj}(a) shows a typical solution for a time-varying conservative delay. The trajectory is characterized by strong fluctuations as known from systems with constant delay and we call the related dynamics \emph{turbulent chaos} in accordance with the term 'optical turbulence' which was introduced in Ref.~\cite{ikeda_optical_1980}. In contrast, for generating the trajectory in Fig.~\ref{fig:traj}(b) only the mean delay was changed from $\tau_0=1.54$ as used for Fig.~\ref{fig:traj}(a) to $\tau_0=1.50$ such that the class of the delay variation changes from conservative to dissipative. The trajectory in Fig.~\ref{fig:traj}(b) is characterized by nearly constant plateaus and burst-like transitions between them. In contrast to the known slowly oscillating periodic solutions for systems with constant delay, the heights of the plateaus during the laminar phases vary chaotically. Since the dynamics is mainly characterized by laminar phases with chaotic intensity variations between these phases, we call this type of chaotic behavior \emph{laminar chaos}. Note that this behavior is also very different from intermittent chaos, which is characterized by laminar phases of fixed intensity, but stochastically varying duration \cite{schuster_deterministic_2005}. Solutions similar to the one presented in Fig.~\ref{fig:traj}(b) can be found for many realizations of Eq.~\eqref{eqn:sysdef} with dissipative delay and large $T$. In the following, we analyze the properties of this hitherto unknown chaotic behavior, its connection to the above mentioned delay classes, and provide the conditions for its appearance.

\begin{figure*}
\centering
Conservative delay \\
\includegraphics[width=0.9\textwidth]{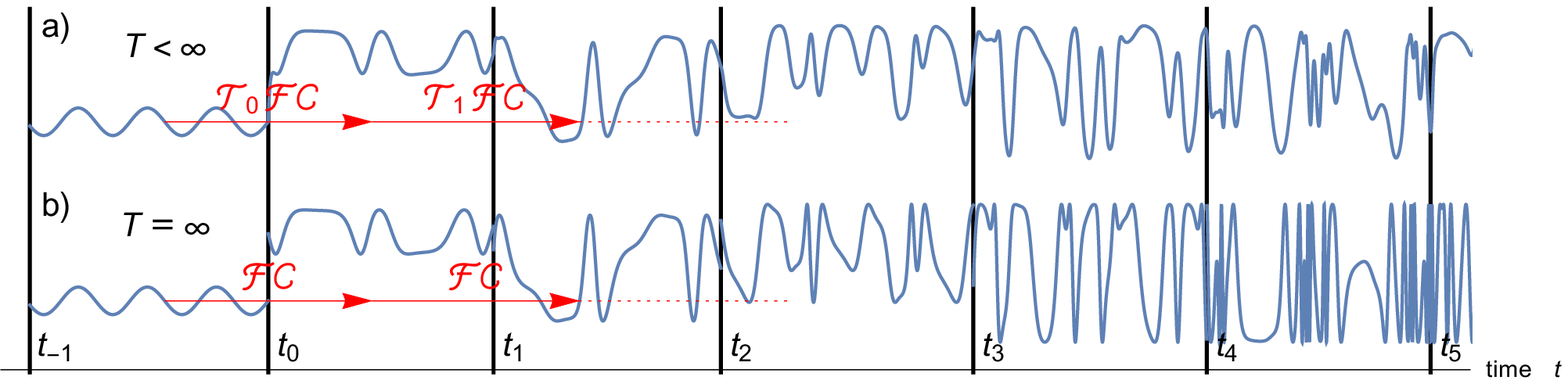} \\
Dissipative delay \\
\includegraphics[width=0.9\textwidth]{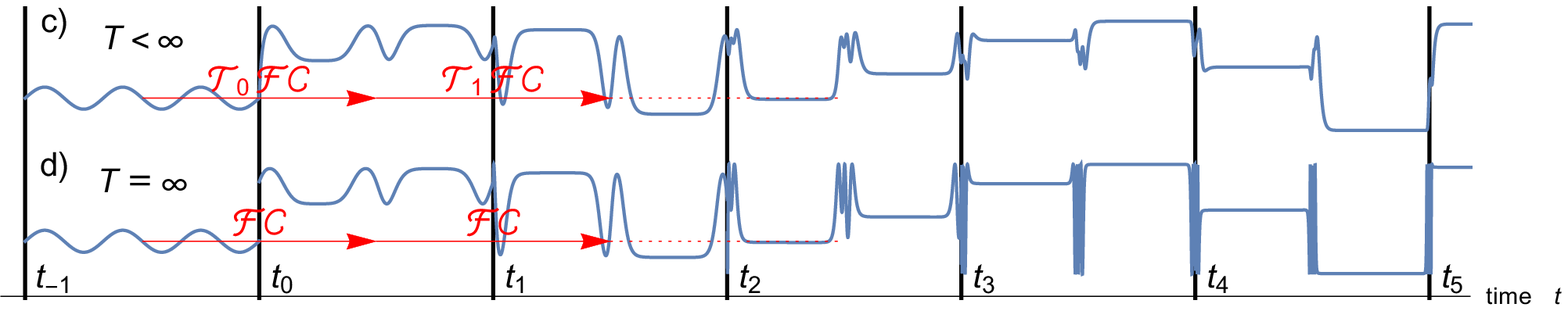} \\
\colorcaption{\label{fig:soluop} Construction of the solution of Eq.~\eqref{eqn:sysdef} for large $T$ via the exact formula Eq.~\eqref{eqn:soluopsep} (a),(c) and the limit map Eq.~\eqref{eqn:limitmap} (b),(d). For conservative delay (a),(b) the operator $\mathcal{C}$ has no significant effect and each iteration of $\mathcal{F}$ creates stronger chaotic oscillations resulting in turbulent chaos. For dissipative delay (c),(d) the periodic repulsive and attracting properties of the map $R^{-1}$ \cite{otto_universal_2017,muller_dynamical_2017} lead to laminar phases and irregular bursts because the chaotic oscillations are compressed at the attracting points of $R^{-1}$. For finite $T$ (a),(c) the additional operator $\mathcal{T}_n$ smooths the solutions compared to the case $T \to \infty$ in (b),(d).}
\end{figure*} 

For the theoretical investigation below we consider periodic delays and we rescale time such that the delay period is equal to one, $\tau(t+1)=\tau(t)$. We further assume the retarded argument $R$ to be invertible, i.e., $\dot{\tau}(t)<1$, and we denote by $R^{-1}$ the inverse. In this case the state intervals of the DDE~\eqref{eqn:sysdef} are given by $(t_{n-1},t_n]$, with $n\in\mathbb{N}$ and $t_{n-1}=R(t_{n})$. We use the \emph{method of steps} for the solution of Eq.~\eqref{eqn:sysdef}, that is, the DDE is integrated stepwise from one state interval to the next  \cite{bellman_computational_1965} and $z_n(t)$ denotes the solution inside the $n$th state interval $t \in (t_{n-1},t_n]$. Starting from the initial function $z_0(t)$, the solution $z_1(t)$ in the first interval can be obtained via the variation of constants formula. In general, the mapping is defined by 
\begin{equation}
\label{eqn:soluop}
z_{n+1}(t) = z_{n}(t_{n}) e^{-T(t-t_{n})} + \int\limits_{t_{n}}^{t} \! dt' \, T e^{-T(t-t')} f(z_n(R(t'))),
\end{equation}
with $t \in (t_{n},t_{n+1}]$. Eq.~\eqref{eqn:soluop} can be divided into three steps, the delay access defined by the Koopman operator $\mathcal{C}$ \cite{koopman_hamiltonian_1931,budisic_applied_2012} with $(\mathcal{C} \psi)(t)=\psi(R(t))$, the nonlinear mapping defined by the operator $\mathcal{F}$ with $(\mathcal{F} \psi)(t)=f(\psi(t))$ and the integration operator $\mathcal{T}_n$, given by $(\mathcal{T}_n \psi)(t)=z(t_n) e^{-T(t-t_n)} + \int_{t_n}^{t} \! dt'\, T e^{-T(t-t')} \psi(t')$. With these definitions Eq.~\eqref{eqn:soluop} can be written as \footnote{A similar decomposition can be found in \cite{otto_universal_2017, *muller_dynamical_2017}, where the integration operator $\mathcal{I}_n$ is defined by $\mathcal{I}_n=\mathcal{T}_n\mathcal{F}$.}
\begin{equation}
\label{eqn:soluopsep}
z_{n}(t) = (\mathcal{T}_{n-1}\mathcal{F}\mathcal{C} z_{n-1})(t).
\end{equation}
For constant delay the dynamics is only characterized by the operators $\mathcal{T}_n$ and $\mathcal{F}$ because in this case $\mathcal{C}$ reduces to a simple shift operator \cite{giacomelli_relationship_1996}. For systems with time-varying delay the Koopman operator $\mathcal{C}$ can have a significant influence on the dynamics of the DDE \cite{otto_universal_2017,muller_dynamical_2017}. The operator $\mathcal{T}_n$ smooths the involved function with the integral kernel $Te^{-T(t-t')}$. For large $T$ the kernel approaches the Dirac $\delta$-distribution and the term $z(t_n) e^{-T(t-t_n)}$ vanishes \cite{ikeda_high-dimensional_1987}. As a consequence, for $T\to\infty$ the operator $\mathcal{T}_n$ becomes the identity and Eq.~\eqref{eqn:soluopsep} simplifies to
\begin{equation}
\label{eqn:limitmap}
z_{n} = \mathcal{F}\mathcal{C} z_{n-1} = f \circ z_{n-1} \circ R = f^n \circ z_{0} \circ R^n,
\end{equation}
which we call \emph{limit map} in extension of the notation used in Refs. \cite{mensour_chaos_1998, larger_flow_2005} for systems with constant delay. Laminar chaos can be found for dissipative delay and large $T$, which means that its main properties can be derived on the basis of Eq.~\eqref{eqn:limitmap}. The dynamics of the limit map can be regarded as the evolution of the graph $(\theta,z_0(\theta))$ representing the initial state of Eq.~\eqref{eqn:limitmap} under iteration of the two-dimensional map $(t',z')=(R^{-1}(t),f(z))$. For $n$ iterations we obtain the parameterized curve 
\begin{equation}
\label{eqn:graph}
(t(\theta),z_n(t(\theta)))=((R^{-1})^n(\theta),f^n(z_0(\theta))),
\end{equation}
where the parameter $\theta$ varies in the initial interval $(t_{-1},t_0]$. The graph $(t,z_n(t))$ represents the state $z_n(t)$ inside the $n$th state interval $(t_{n-1},t_n]$ corresponding to the $n$-fold application of Eq.~\eqref{eqn:limitmap} to the initial state $z_0(\theta)$. Obviously, the two-dimensional map decomposes into two independent one-dimensional maps
\begin{subequations}
\begin{align}
x_n &= R^{-1}(x_{n-1}) = R^{-n}(x_0), \label{eqn:2dmapb}\\
y_n &= f(y_{n-1}) = f^n(y_0). \label{eqn:2dmapa}
\end{align}
\end{subequations}
The map $R^{-1}$ specifies the position $x_n$ on the $x$-axis corresponding to the time axis of the DDE and the map $f$ or equivalently the operator $\mathcal{F}$ creates the function values $y_n$ at the points $x_n$. 

\begin{figure}[t]
\includegraphics[width=0.4\textwidth]{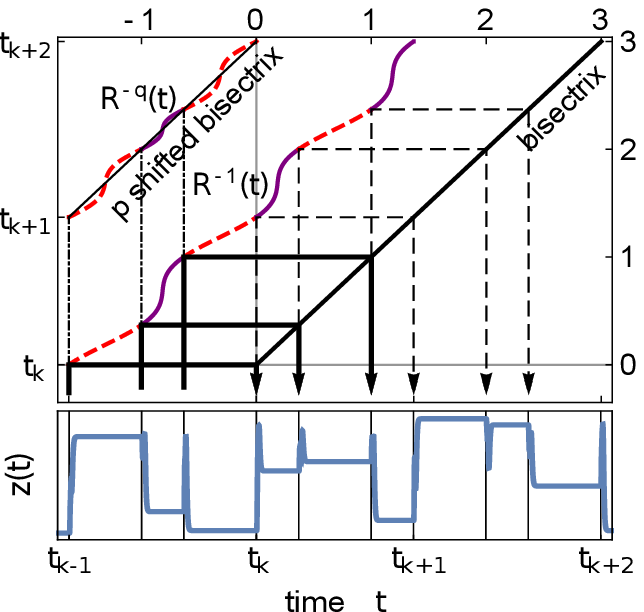}
\colorcaption{\label{fig:qperiosep} A laminar chaotic solution $z(t)$ and the inverse access map Eq.~\eqref{eqn:2dmapb} of the corresponding dissipative delay with rotation number $\rho=\frac{p}{q}$ are shown (here $\rho=\frac{3}{2}$). The arrows indicate the mapping of the boundaries of the laminar phases to the next state interval, which correspond to the attractive periodic points of $R^{-1}\mod 1$ (the attractive fixed points of $R^{-q}(t)-p$, where the subtraction of $p$ removes the drift). Thus, the width and the location of the plateaus is specified by the variable delay $\tau(t)$, whereas its intensity levels are given by the nonlinearity $f$ (see text).}
\end{figure}

The construction of the solution of the DDE~\eqref{eqn:sysdef} for conservative and dissipative delays with the exact formula Eq.~\eqref{eqn:soluopsep} and with the limit map Eq.~\eqref{eqn:limitmap} associated with Eqs.~\eqref{eqn:2dmapb} and \eqref{eqn:2dmapa} is illustrated in Fig.~\ref{fig:soluop}. For conservative delay the reduced map $R^{-1}\mod 1$ exhibits quasiperiodic dynamics which preserves the mean distance between the points $x_n$. If $f$ exhibits chaotic dynamics, due to the sensitivity on initial conditions, variations in $(\theta,z_0(\theta))$ cause oscillations in $(t,z_n(t))$, which get stronger for increasing $n$ [see Fig.~\ref{fig:soluop}(b)]. For finite $T$ the additional smoothing operator $\mathcal{T}_n$ damps high frequencies \cite{ikeda_high-dimensional_1987,giacomelli_relationship_1996}. Thus, for conservative delay the operator $\mathcal{C}$ has no significant influence on the dynamics (cf. \cite{otto_universal_2017,muller_dynamical_2017}) and turbulent chaos appears as known from systems with constant delay [see Fig.~\ref{fig:soluop}(a)]. In contrast, for dissipative delay the dynamics under iterations of the reduced map $R^{-1}\mod 1$ is characterized by mode-locking and attracting motion with rational rotation number $\rho=\frac{p}{q}$ (cf. Refs.~\cite{katok_introduction_1997, ott_chaos_2002}). There are $p$ stable equilibria or stable periodic points inside each state interval $(t_{n-1},t_{n}]$ and the $x_n$ accumulate at these attracting points under iterations of Eq.~\eqref{eqn:2dmapb}. Consequently, the graph $(t,z_n(t))$ develops $p$ plateaus separated by the attracting points where the oscillations from the iterations of the map $f$ accumulate [see Fig.~\ref{fig:soluop}(d)]. Similar to the case of a conservative delay, for finite $T$, the smoothing operator $\mathcal{T}_n$ in Eq.~\eqref{eqn:soluopsep} damps high frequency oscillations, which appear, for dissipative delays, only at the plateau boundaries of the laminar chaotic solution [see Fig.~\ref{fig:soluop}(c)]. Thus, the fundamentally different properties of the operator $\mathcal{C}$ for dissipative delays facilitate the existence of laminar chaos. As illustrated in Fig.~\ref{fig:qperiosep}, each of the $p$ plateaus inside the state interval $(t_{n-1},t_{n}]$ is mapped uniquely to one plateau inside the next state interval $(t_{n},t_{n+1}]$. The width of the laminar phases (plateaus) depends on the arrangement of the attracting points of the map $R^{-1}$, whereas the intensity levels $y_n$ during the laminar phases in the $n$th state interval are connected to the intensity levels $y_{n-1}$ of the corresponding laminar phases in the previous state interval via the function $f$ as in Eq.~\eqref{eqn:2dmapa}. Conversely, the latter property allows for a simple determination of the nonlinearity $f$ of the delay equation, Eq.~\eqref{eqn:sysdef}, by experimentally observing the intensity level variation of laminar chaos.

In the following we derive a quantitative criterion for the existence of plateaus in the limit map Eq.~\eqref{eqn:limitmap}, which is a necessary condition for the existence of laminar chaos in systems with dissipative delay. Plateaus exist if the time derivative $\dot{z}_n(t)$ vanishes between the plateau boundaries, i.e., between the attracting points of the map $R^{-1}$. The derivative of $z_n(t)$ is obtained via Eq.~\eqref{eqn:limitmap} as
\begin{equation}
\label{eqn:limitslope}
\dot{z}_n(t)  = \left. (f^n)'(y) \right\vert_{y=z_0(R^n(t))} \left.z'_0(\theta) \right\vert_{\theta=R^n(t)} (R^n)'(t).
\end{equation}
For large $n$, $R^n(t)$ converges to $\theta^*$, where $\theta^*$ is an attracting point of the map $R \mod 1$, that is a repulsive point of $R^{-1}$ located within the plateaus (cf. Fig.~\ref{fig:qperiosep}), and correspondingly $z'_0(R^n(t)) \to z'_0(\theta^*)=\text{const.}$ The derivatives of $f^n$ and $R^n$ increase or decrease exponentially for large $n$, $\vert (f^n)'(y) \vert \sim e^{n \lambda[f]}$ and $\vert (R^n)'(t) \vert \sim e^{n \lambda[R]}$, where $\lambda[f]$ and $\lambda[R]$ denotes the Lyapunov exponent (cf. Ref.~\cite{katok_introduction_1997}) of the maps $f$ and $R$, respectively. As a result, for increasing $n$ the derivative $\dot{z}_n(t)$ converges to zero and the solution $z_n(t)$ of the limit map Eq.~\eqref{eqn:limitmap} becomes constant between two attracting points of the map $R^{-1}$ if 
\begin{equation}
\label{eqn:lamcrit}
\lambda[f] + \lambda[R] < 0.
\end{equation}
Thus, laminar chaos can be observed, only if Eq.~\eqref{eqn:lamcrit} holds and $\lambda[f]>0$.
Eq.~\eqref{eqn:lamcrit} holds exactly in the limit $T \to \infty$. For finite $T$ the width of the integration kernel of the additional smoothing operator $\mathcal{T}_n$ and therefore the width of the irregular bursts at the boundaries of the laminar phases scales with $1/T$. As a consequence, laminar phases can develop only if the distance between two attracting points of the map $R^{-1}$ is sufficiently larger than $1/T$.


\begin{figure}[t]
\includegraphics[width=0.42\textwidth]{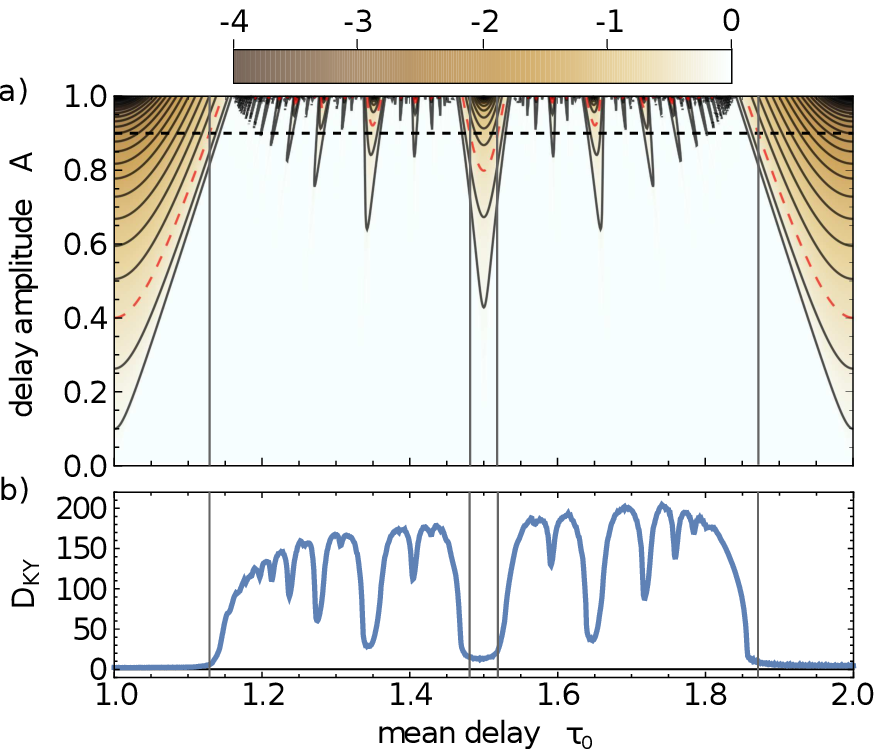}
\colorcaption{\label{fig:dim} a) Lyapunov chart of the access map $R$ for $\tau(t)=\tau_0+A\sin(2\pi\,t)/(2\pi)$, where the contours correspond to a constant Lyapunov exponent $\lambda[R](\tau_0,A)$. For Eq.~\eqref{eqn:sysdef} with $f(z)=2\,z/(1+z^{10})$ (Mackey-Glass equation) the criterion~\eqref{eqn:lamcrit} for laminar chaos is fulfilled above the dashed (red) line, which in this case corresponds to the contour level $-\lambda[f] \approx -0.51$. b) Kaplan-Yorke dimension $D_{KY}$ of this system ($T=2000$) for the sinusoidal delay with $A=0.9$ (horizontal dashed line in (a)). For conservative delays very high attractor dimensions are possible. In contrast, for mean delays $\tau_0$, where Eq.~\eqref{eqn:lamcrit} holds, $D_{KY}$ is very small indicating laminar chaos.}
\end{figure}

The occurrence of laminar chaos can depend sensitively on the parameters of the delay $\tau(t)$ because, for example, the Lyapunov exponent of the access map $R$ changes in a fractal manner with the parameters (cf. \cite{otto_universal_2017,muller_dynamical_2017}). The latter is well-known for circle maps and can be illustrated by the Lyapunov chart $\lambda[R](\tau_0,A)$  \cite{de_figueiredo_lyapunov_1998} illustrated in Fig.~\ref{fig:dim}(a) for an exemplary access map. The criterion Eq.~\eqref{eqn:lamcrit} holds above the contour $\lambda[R](\tau_0,A)=-\lambda[f]$ [see dashed red line in Fig.~\ref{fig:dim}(a)]. For $A=1$ there is one superstable orbit ($\lambda[R]=-\infty$) inside each mode-locking regime \cite{glass_fine_1982}. Thus, for each dissipative delay $\tau(t)$ and each rational rotation number there exists a finite region in parameter space, namely in a neighborhood of the superstable parameter points, where laminar chaos is possible. In Fig.~\ref{fig:dim}(b) the Kaplan-Yorke dimension \cite{ott_chaos_2002} of the Mackey-Glass equation with time-varying delay for large $T$ is plotted for fixed amplitude $A=0.9$ as a function of the mean delay $\tau_0$. For conservative delay we have $\lambda[R]=0$. Consequently, turbulent chaos appears, which is characterized by a large Kaplan-Yorke dimension. In contrast, for dissipative delays with $\lambda[f]+\lambda[R](\tau_0,A)<0$ laminar chaos with a very low Kaplan-Yorke dimension appears. One can also observe that there exist in addition dissipative delays with $\lambda[R](\tau_0,A)<0$, where the criterion \eqref{eqn:lamcrit} is not fulfilled but other local minima of the Kaplan-Yorke dimension can be observed. A study of these intermediate states, which are qualitatively different from laminar and turbulent chaos, will be presented elsewhere.

In conclusion, we have presented a hitherto unknown type of chaotic behavior, which we call laminar chaos. It is characterized by laminar phases with chaotically varying intensity and burst-like transitions between them. Laminar chaos is observed, for example, in systems with large dissipative delays because for large delays the influence of the integration operator $\mathcal{T}_n$ vanishes, whereas the properties of dissipative delays become more important. In the present Letter, laminar chaos was only studied for scalar DDEs. However, similar phenomena exist in more general nonscalar systems, which we observed, for example, in the Lang-Kobayashi equations. For higher dimensional maps $f$, one can obtain laminar chaos in one direction and turbulent chaos in another direction. Moreover, for more complex access maps $R$ it may be possible to obtain a temporal switching between laminar and turbulent phases. Laminar chaotic solutions may also exist in other systems such as renewal equations with dissipative delay. The shape of the chaotic solutions can be tuned by changing the properties of the retarded argument $R$ and the nonlinearity $f$, which might be interesting for applications. Since dynamical systems with time-varying delay can be realized by optoelectronic experiments, the presented dynamical behavior, in principle, can be observed experimentally and may lead to new applications or improvements of existent applications. For example, in information processing technologies such as chaos communication \cite{goedgebuer_optical_1998, *vanwiggeren_optical_1998, *udaltsov_communicating_2001, *keuninckx_encryption_2017} and reservoir computing \cite{appeltant_information_2011, *larger_high-speed_2017} the laminar phases may be used as information units, where their intensity levels code the information to be processed. 

We acknowledge partial support from the German Research Foundation (DFG) under the grant no. 321138034.
\bibliography{ddelamchaos}

\end{document}